 \documentstyle[aps,preprint]{revtex}

\draft

\begin{document}
\title{Coherent population trapping in two-electron three-level systems with
aligned spins}
\author{Jordi Mompart $^1$, Ramon Corbalan $^1$, and Luis Roso $^{1,2}$}
\address{$^1$ Departament de F\'\i sica, Universitat Aut\`onoma de\\
Barcelona, E-08193 Bellaterra, Spain}
\address{$^2$
Departamento de Fisica Aplicada,\\
Universidad de Salamanca, E-37008 Salamanca, Spain}
\maketitle

\begin{abstract}
The possibility of coherent population trapping in two electron states with
aligned spins ({\it ortho-system}) is evidenced. From the analysis of a
three-level atomic system containing two electrons, and driven by the two
laser fields needed for coherent population trapping, a conceptually new
kind of two-electron dark state appears. The properties of this trapping are
studied and are physically interpreted in terms of a dark hole, instead of a
dark two-electron state. This technique, among many other applications,
offers the possibility of measuring, with subnatural resolution, 
some superposition-state matrix-elements 
of the electron-electron correlation that due to their time
dependent nature are inaccesible by standard measuring procedures.

\end{abstract}



Coherent manipulation of the atomic wavefunction by laser fields allows the
preparation of the atom in particular states with surprising properties due
to the constructive or destructive interference of the wavefunction. The
presence of non-absorption resonances or dark states is well known since the
late seventies \cite{alzetta} \cite{orriols} \cite{stroud}. From those
initial times the physical properties of the dark and bright states have
been fruitfully employed in thousands of different situations \cite{arimondo}.
Related concepts such as Electromagnetically Induced Transparency, Amplification
Without Inversion, or Lasing Without Inversion, have been also introduced over the
last years \cite{harris} \cite{mompart}.

Most of the related experiments deal with single electron atoms, or single
active electron situations, where an electron is coherently forced to a
superposition state. In a two electron system, however, the situation can be
a bit different because of the electron-electron interaction and
particularly because of the Pauli exclusion principle. If both electrons are
not allowed to be at the same state at the same time, the dynamics of one of
them will strongly influence the dynamics of the second one. Of course,  to
evidence this effect one needs to work with electrons having parallel spins. If
spins were antiparallel, then both of them could be allowed in the same atomic
(spatial) state. There are some studies that use related properties of the two
electron systems in case of antiparallel spins
\cite{grobe}, 
in different contexts, namely double-core resonance in two-electron atoms.

In the present letter we restrict ourselves to the case of a two-electron
{\it ortho-system} \cite{orto-para}, 
i.e. a case in which both electrons have parallel spins and
therefore the Pauli exclusion principle acts on the spatial part of the
wavefunctions. This prevents both electrons to be in the same spatial state. 
For the case of coherent population trapping this has rich consequences of a very
fundamental nature that have never been considered before. We present the
general case, we discuss some possibilities for experimental realization of
such a system, and we propose the use of coherent population trapping to measure
some time-dependent electron-electron interaction matrix elements.

Let us consider a typical V-configuration of the three single-electron states,
as indicated in Fig. 1,
labelled $\vert a \rangle$, $\vert b \rangle$, and $\vert c \rangle$,
with energies  
$\hbar \omega_a$, $\hbar \omega_b$, and $\hbar \omega_c$ respectively. 
If two of these states are populated, we can build
up antisymmetrized two-electron states. Since two electrons with aligned
spins are considered, the spin term is symmetric and the antisymmetry of the
total wave function comes from the spatial part of the wavefunction.

The three level system is formed by two adjacent dipole transitions sharing
a common state $\vert c \rangle$. 
The dipole coupling to the laser fields, 
$H_{dip}$, is given by 
$\langle a \vert H_{dip} \vert c \rangle =\hbar\alpha e^{-i\omega_\alpha t} $, and 
$\langle b \vert H_{dip} \vert c \rangle =\hbar\beta e^{-i\omega_\beta t} $, where
$\alpha$ and $\beta$ are the Rabi frequencies of the two laser fields, and 
$\omega_\alpha$ and $\omega_\beta$ represent their frequencies.
We assume
that all three single-electron states involved have definite parity, with states 
$\vert a \rangle$ and 
$\vert b \rangle$ having the same parity and state 
$\vert c \rangle$ having the opposite one. 
Therefore, 
$\langle a \vert H_{dip} \vert a \rangle =$ 
$\langle b \vert H_{dip} \vert b \rangle = $ 
$\langle c \vert H_{dip} \vert c \rangle = 0 $, and also 
$\langle a \vert H_{dip} \vert b \rangle = 0 $.
We introduce the Rotating Wave Approximation \cite{orriols} \cite{rwa}, 
just keeping slow oscillations at frequencies comparable to the detunings, 
$\Delta_{\alpha}= \omega_\alpha - (\omega_a-\omega_c) $, and 
$\Delta_{\beta}= \omega_\beta - (\omega_b -\omega_c) $. 
To describe the long-time dynamics of
this system is necessary to consider relaxation.
We introduce the relaxation coefficient $\gamma_{ac}$, indicating the rate
of decay of the population from state $\vert a \rangle$ (the uppermost
level) to state $\vert c \rangle$, and the relaxation coefficient 
$\gamma_{bc}$, indicating the rate of decay of the population from state 
$\vert b \rangle$ to state $\vert c \rangle$. In the dipole approximation, 
$\gamma_{ba}=\gamma_{ab}=0$

Let us consider now that two electrons (in the same spin state) are forced
to be inside this three-level system.  The two-electron states are indicated
by $\vert i, j \rangle =\vert i \rangle \bigotimes \vert j \rangle$, with $i,j= a,b,c$.
Antisymmetrized states are $\vert A \rangle = {\frac{1 }{\sqrt 2}} \left(
\vert c, b \rangle -\vert b, c \rangle \right) $, $\vert B \rangle={\frac{1 
}{\sqrt 2}} \left( \vert a, c \rangle -\vert c, a \rangle \right)$, and $%
\vert C \rangle={\frac{1 }{\sqrt 2}} \left( \vert a, b \rangle -\vert b, a
\rangle \right)$. Without considering electron-electron interaction, the
energies of those states are given by $\hbar$ times $\omega_A = \omega_b+
\omega_c$, for state $\vert A \rangle$, $\omega_B = \omega_a+ \omega_c$, 
for
state $\vert B \rangle$, and $\omega_C = \omega_a+ \omega_b$, for state 
$\vert C \rangle$.

Parity of the two-electron states is directly related to the parity of the
single-electron states. Two electron states will thus keep a well defined
parity. In particular, $\vert A\rangle $ and $\vert B\rangle $ will be of the same
parity and $\vert C\rangle $ of the opposite one. 
The two-electron states non-vanishing dipole matrix elements will be: 
$\langle C\vert H_{dip}\vert A\rangle =\hbar \alpha e^{-i\omega _{\alpha }t}$, and 
$\langle C\vert H_{dip}\vert B\rangle =\hbar \beta e^{-i\omega _{\beta }t}$ 
and all other couplings
are zero because of parity considerations, 
$\langle A\vert H_{dip}\vert A\rangle =$ 
$\langle B\vert H_{dip}\vert B\rangle =$
$\langle C\vert H_{dip}\vert C\rangle =$
$\langle A\vert H_{dip}\vert B\rangle =0$. 
These two-electron states determine now a $\Lambda
$-configuration, as indicated in Fig. 2, while the one-electron states $\vert a\rangle $,
$\vert b\rangle $, $ \vert c\rangle $ formed a V-configuration.

The density matrix for the two-electron three-level system is defined by 
$ (\rho )_{IJ}=\vert I\rangle \langle J\vert $, with $I,J=A,B,C$. 
Still without
considering the electron-electron interaction, the dynamical equations for
the two-electron density matrix are, 
\begin{mathletters}
\begin{eqnarray}
{\frac{d}{dt}}\rho _{AA} &=&i\left[ \rho _{AC}\alpha -\rho
_{AC}^{\ast }\alpha^{\ast } \right] +\gamma _{CA}\rho _{CC} \\
{\frac{d}{dt}}\rho _{BB} &=&i\left[ \rho _{BC}\beta -\rho
_{BC}^{\ast }\beta^{\ast } \right] +\gamma _{CB}\rho _{CC} \\
{\frac{d}{dt}}\rho _{CC} &=&i\left[ -\rho _{AC}\alpha +\rho
_{AC}^{\ast }\alpha^{\ast } -\rho _{BC}\beta +\rho _{BC}^{\ast }\beta^{\ast }
\right] -\left( \gamma _{CA}+\gamma _{CB}\right) \rho _{CC} \\
{\frac{d}{dt}}\rho _{AB} &=&i\left[ -\rho _{AC}\left( \Delta _{\alpha
}-\Delta _{\beta }\right) +\rho _{AC}\beta -\rho _{BC}^{\ast }\alpha^{\ast }
\right] -\Gamma _{AB}\rho _{AB} \\
{\frac{d}{dt}}\rho _{AC} &=&i\left[ -\rho _{AC}\Delta _{\alpha }+\alpha^{\ast }
\left( \rho _{AA}-\rho _{CC}\right) +\rho _{AB}\beta^{\ast } \right] -\Gamma
_{AC}\rho _{AC} \\
{\frac{d}{dt}}\rho _{BC} &=&i\left[ -\rho _{BC}\Delta _{\beta }+\beta^{\ast } 
\left( \rho _{BB}-\rho _{CC}\right) +\rho _{AB}^{\ast }\alpha^{\ast } \right]
-\Gamma _{BC}\rho _{BC}
\end{eqnarray}
The relaxations correspond to single electron processes, $\gamma
_{CA}=\gamma _{ac}$, $\gamma _{CB}=\gamma _{bc}$, indicate an electron
falling from state $\vert a\rangle $ to state $\vert c\rangle $, and from state $%
\vert b\rangle $ to state $\vert c\rangle $, respectively. Dipole decay between states 
$\vert a\rangle $ and $\vert b\rangle $ is forbidden by parity: $\gamma _{BA}=\gamma
_{ab}=0$, and $\gamma _{AB}=\gamma _{ba}=0$. The two-electron coherences
will relax according to: $2\Gamma _{BA}=\gamma _{BA}+\gamma _{AB}=0$, 
$ 2\Gamma _{AC}=\gamma _{CA}+\gamma _{CB}=\gamma _{ac}+\gamma _{bc}$, 
and 
$2\Gamma _{BC}=\gamma _{CA}+\gamma _{CB}=\gamma _{ac}+\gamma _{bc}$. 
No simultaneous two-electron relaxation mechanisms will be considered, because
they are forbidden in the electric dipole approximation.

Following standard ideas of solid state physics we can interpret that we
have two electrons to fill three states, 
$\vert a\rangle $, $\vert b\rangle $, and $\vert c\rangle $. 
So there are two occupied states plus an empty one, the hole.
The system is thus characterized by the position of the hole. The
two-electron state $\vert C\rangle $, for example, involves one electron at state 
$\vert a\rangle $ and the second electron at state $\vert b\rangle $, leaving the 
$\vert c\rangle $ empty. 

It is perfectly well established  \cite{alzetta} \cite{orriols} \cite{stroud}, 
that under certain conditions the two lower states form a very
peculiar coherent superposition. 
The result is that a state $\vert d\rangle $
uncoupled to the laser fields and another state $\vert e\rangle $ coupled to the
laser fields appear. 
These single-electron states are not precisely the
typical dark and bright states because they correspond to the two upper
states of the V-configuration and relax very fast. In the particular case
that the Rabi frequencies of both transitions are equal, $\alpha =\beta $,
and the detunings verify $\Delta _{\alpha }=\Delta _{\beta }$, then the
expressions of the dark, $\vert d\rangle $, and the bright, $\vert e\rangle $, states
are extremely simple and symmetrical, $\vert d\rangle ={\frac{1}{\sqrt{2}}}\left(
\vert a\rangle -\vert b\rangle \right) $, and $\vert e\rangle ={\frac{1}{\sqrt{2}}}\left(
\vert a\rangle +\vert b\rangle \right) $.

We can repeat the procedure to obtain the dark states now using two-electron
states. The result is that a dark state $\vert D\rangle $ and a bright state 
$\vert E\rangle $ appear. 
In the particular case of equal Rabi frequencies, $\alpha =\beta $, 
and equal detunings $\Delta _{\alpha }=\Delta _{\beta }$, 
then the dark state is 
$\vert D\rangle={\frac{1}{\sqrt{2}}}\left( \vert A\rangle -\vert B\rangle \right)$ 
and the bright state is
$\vert E\rangle ={\frac{1}{\sqrt{2}}}\left( \vert A\rangle +\vert B\rangle \right)$.  
Now the dark state $\vert D\rangle$ 
involves one electron in a superposition of the upper states 
$\vert a\rangle $ and $\vert b\rangle $, 
while the other electron lies in  the lowest energy state $\vert c\rangle $. 
Therefore a hole appears in the empty upper state. 
Relaxation does not play a role now because the
lower state is filled with an electron and Pauli exclusion principle does not allow a
second one in the same state. 
Therefore this dark state has the appearance of a hole
that is moving between the two upper single electron states. 
To illustrate this we have included a new scheme of the single electron states. 
Fig. 3a correspond to only one electron (grey circle) in the system. 
Under the appropriated conditions, the
electron can be placed in a state $\vert d\rangle $ that is not coupled to the fields.
This state, however, can decay. In the case of two electrons, Fig. 3b, there is one
electron at the lower state that prevents the decay from the upper states. The state
without electron, the hole, is trapped and stable. 
Therefore,  we have now {\it a hole placed at a dark state!}

We have so far considered two-electron systems where the electron-electron
interaction is not accounted for. Only the Pauli exclusion principle has
been considered through the antisymmetrization of the spatial part of the
wave function (spin part is always symmetrical in the considered {\it %
ortho-atom}). The two-electron system can be understood in terms of a hole.
Of course this agreement is so far perfect because we have forgotten one
particular feature, the electron-electron repulsion. Now it is time to
consider these terms and see how they modify the presented results. The
interaction Hamiltonian is $H_{ee}=e^{2}/\vert r_{1}-r_{2}\vert $, $r_{1}$ and $r_{2}$
being the positions of the electrons, then 
\end{mathletters}
\begin{mathletters}
\begin{eqnarray}
\langle A\vert H_{ee}\vert A\rangle  &=&\langle cb\vert H_{ee}\vert cb\rangle -\langle
cb\vert H_{ee}\vert bc\rangle =\hbar \Delta _{A} \\
\langle B\vert H_{ee}\vert B\rangle  &=&\langle ca\vert H_{ee}\vert ca\rangle -\langle
ca\vert H_{ee}\vert ac\rangle =\hbar \Delta _{B} \\
\langle C\vert H_{ee}\vert C\rangle  &=&\langle ba\vert H_{ee}\vert ba\rangle -\langle
ba\vert H_{ee}\vert ab\rangle =\hbar \Delta _{C} \\
\langle A\vert H_{ee}\vert C\rangle  &=&\langle cb\vert H_{ee}\vert ab\rangle -\langle
cb\vert H_{ee}\vert ba\rangle =0 \\
\langle B\vert H_{ee}\vert C\rangle  &=&\langle ca\vert H_{ee}\vert ab\rangle -\langle
ca\vert H_{ee}\vert bc\rangle =0 \\
\langle A\vert H_{ee}\vert B\rangle  &=&\langle cb\vert H_{ee}\vert ac\rangle -\langle
cb\vert H_{ee}\vert ca\rangle =\hbar \chi e^{-i\omega _{b}t}e^{i\omega _{a}t}
\end{eqnarray}
These electron-electron interaction terms can be grouped in two different
families, the time-independent terms (that do not contain time oscillations
at the energy difference) and the time-dependent terms (that do contain
explicit time oscillations). Time independent terms come from the diagonal
matrix elements 
$\langle A\vert H_{ee}\vert A\rangle $, 
$\langle B\vert H_{ee}\vert B\rangle $, 
$\langle C\vert H_{ee}\vert C\rangle $. 
The contributions 
$\langle c,b\vert H_{ee}\vert c,b\rangle$, 
$\langle c,a\vert H_{ee}\vert c,a\rangle $, and
$\langle b,a\vert H_{ee}\vert b,a\rangle $, are Coulomb terms. 
The contributions
$\langle c,b\vert H_{ee}\vert b,c\rangle $, 
$\langle c,a\vert H_{ee}\vert a,c\rangle $, and 
$\langle b,a\vert H_{ee}\vert a,b\rangle $, are exchange terms. 
For a system in a pure quantum state only these two kind of
electron-electron terms are relevant. With the dynamical situation
established via the interaction with the laser fields, electrons are in
superposition states and, thus, new terms may appear that are time
dependent. The term, $\langle A\vert H_{ee}\vert B\rangle $ is very particular because
it involves three different one-electron states. It is the sum of two
different contributions 
$\langle c,b\vert H_{ee}\vert a,c\rangle $, and 
$\langle c,b\vert H_{ee}\vert c,a\rangle $. 
The cross term $\langle cb\vert H_{ee}\vert ac\rangle $  
involves the two dipole-allowed transitions, 
$\vert a\rangle \leftrightarrow \vert c\rangle$ for one electron, and 
$\vert c\rangle \leftrightarrow \vert b\rangle$ for the other electron. 
Another cross term that appears is 
$\langle c,b\vert H_{ee}\vert c,a\rangle $ , 
it is different to the first one because  one of the electrons 
remains in the common state $c$ while 
the other is in a  
$\vert a\rangle \leftrightarrow \vert b\rangle$ coherence. 
In any case, these two terms present a time oscillation at frequency
$\omega _{a}-\omega _{b}$, 
so the resulting electron-electron matrix element can be
written as, 
$\langle A\vert H_{ee}\vert B\rangle =
\hbar \chi e^{-i\omega_{b}t}e^{i\omega _{a}t}$. 
Finally, let us remember that due to parity considerations,
$\langle A\vert H_{ee}\vert C\rangle = \langle B\vert H_{ee}\vert C\rangle =0$ 
because  $H_{ee}$ is an even operator.

Now we can introduce the $H_{ee}$ couplings in the time evolution of the
density matrix. To simplify the expressions it is worth to consider as the
energy origin the corrected energy of state $\vert A\rangle $. After introducing
the rotating wave approximation, we obtain a new system of dynamical
equations 
\end{mathletters}
\begin{mathletters}
\begin{eqnarray}
{\frac{d}{dt}}\rho _{AA} &=&i\left[ \rho _{AC}\alpha -\rho
_{AC}^{\ast }\alpha^{\ast } \right] +\gamma _{CA}\rho _{CC} \\
&+&i\left[ \rho _{AB}\chi ^{\ast }e^{i\omega _{a}t}e^{-i\omega _{b}t}-\rho
_{AB}^{\ast }\chi e^{-i\omega _{a}t}e^{i\omega _{b}t}\right]   \nonumber \\
{\frac{d}{dt}}\rho _{BB} &=&i\left[ \rho _{BC}\beta -\rho
_{BC}^{\ast }\beta^{\ast } \right] +\gamma _{CB}\rho _{CC} \\
&+&i\left[ \rho _{AB}^{\ast }\chi e^{-i\omega _{a}t}e^{i\omega _{b}t}-\rho
_{AB}\chi ^{\ast }e^{i\omega _{a}t}e^{-i\omega _{b}t}\right]   \nonumber \\
{\frac{d}{dt}}\rho _{CC} &=&i\left[ -\rho _{AC}\alpha +\rho
_{AC}^{\ast }\alpha^{\ast } -\rho _{BC}\beta +\rho _{BC}^{\ast }\beta^{\ast } \right]
-\left( \gamma _{CA}+\gamma _{CB}\right) \rho _{CC} \\
{\frac{d}{dt}}\rho _{AB} &=&i\left[ -\rho _{AC}\left( \Delta _{\alpha
}-\Delta _{\beta }\right) +\rho _{AC}\beta -\rho _{BC}^{\ast }\alpha^{\ast }
\right] -\Gamma _{AB}\rho _{AB} \\
&+&i\left[ \left( \rho _{AA}-\rho _{BB}\right) \chi e^{-i\omega
_{a}t}e^{i\omega _{b}t}+\rho _{AB}\left( \Delta _{A}-\Delta _{B}\right) 
\right]   \nonumber \\
{\frac{d}{dt}}\rho _{AC} &=&i\left[ -\rho _{AC}\Delta _{\alpha }+\alpha^{\ast }
\left( \rho _{AA}-\rho _{CC}\right) +\rho _{AB}\beta^{\ast } \right] -\Gamma
_{AC}\rho _{AC} \\
&+&i\left[ \rho _{AC}\left( \Delta _{C}-\Delta _{A}\right) -\rho _{BC}\chi
e^{-i\omega _{\alpha }t}e^{i\omega _{\beta }t}\right]   \nonumber \\
{\frac{d}{dt}}\rho _{BC} 
&=&
i\left[ -\rho _{BC}\Delta _{\beta }+\beta^{\ast } \left( \rho _{BB}-\rho _{CC}\right) 
+\rho _{AB}^{\ast }\alpha^{\ast } \right] -\Gamma_{BC}\rho _{BC} \\
&+&
i\left[ \rho _{BC}\left( \Delta _{C}-\Delta _{A}\right) 
-\rho _{AC}\chi^{\ast }e^{i\omega _{\alpha }t}e^{-i\omega _{\beta }t}\right]   
\nonumber
\end{eqnarray}

One very clear example where coherent population trapping of this kind is
calcium, or other alkaline-earth elements. 
In the ortho-atom case (parallel spins) 
the ground state is of the form n$s$n$p$ (4$s$4$p$ for calcium). 
One excited state very interesting for our purposes is the n$p$n$p$. 
The transition energy is about 2.9 eV for calcium. 
If a $\sigma ^{+}$, $\sigma ^{-}$ laser field is considered, 
then the three single electron states involved will be 
$\vert a\rangle =\vert {\rm n}p_{+}\rangle $, 
$\vert b\rangle =\vert {\rm n}p_{-}\rangle $, and 
$\vert c\rangle =\vert {\rm n}s \rangle $. 
Obviously, the three-two electron states will become 
$\vert A\rangle =\vert {\rm n}s,{\rm n}p_{-}\rangle $, 
$\vert B\rangle =\vert {\rm n}s,{\rm n}p_{+}\rangle $, and 
$\vert C\rangle =\vert {\rm n}p_{+},{\rm n} p_{-}\rangle $. 
In this particular case, $\langle A\vert H_{ee}\vert B\rangle =0$ due
to the angular momentum addition laws \cite{angular}. 
Because n$s$n$p_{+}$ and n$s$n$p_{-}$ 
correspond to the ground state of the ortho-atom, they are
relatively stable. In that case a clear dark state can be formed with all
the characteristics of the coherent population trapping, except that the
trapping affects the hole, that is in a coherent superposition of the two
excited single-electron states. Other similar systems can be found in atoms
with more than two electrons in the external shell. 
This structure is also present in ortho-lithium (lithium with three aligned spins) 
involving the 1$s$2$s$2$p$ and 1$s$2$p$2$p$ states. 

On the opposite case, for atoms or molecules where the $\langle A \vert
H_{ee} \vert B \rangle$ matrix element is non-zero, we predict a similar
dark resonance but shifted by an amount equal to the time average of the 
$\langle A \vert H_{ee} \vert B \rangle$ coupling. 
This suggest a new method to calculate electron-electron correlations 
with subnatural resolution. 
To illustrate this, Fig. 4 has been included. 
This figure represents the spectra of the dark state resonance and its two satellites induced by the
crossed correlations terms of the electron-electron interaction. 
It is a plot the imaginary part of the 
$\vert A \rangle \leftrightarrow \vert C\rangle$
coherence versus the $\Delta_{\alpha}$ detuning,
and corresponds to 
$\gamma_{ac}=\gamma_{bc}$, 
$\alpha=0.1 \gamma_{bc}$, 
$\beta=0.1 \gamma_{bc}$, 
$\Delta_{\beta}=0$, and 
$\chi=0.3 \gamma_{bc}$.
For simplicity, all parameters have been given referred to $\gamma_{bc}$.
 
Coherent population trapping appears at the center 
$\Delta_{\alpha}=0$, as expected.
Moreover, two satellite holes appear at $\pm \chi$. They are due to the
splitting of the levels induced by the time dependent matrix element and to
its coupling to the laser field. The dresed transitions responsible for those peaks are
depicted at the insets of Fig. 4.

In conclusion we have analyzed the dynamical properties of a three-level
two-electron system. The presence of a coherence between the atomic
two-electron states leads to population trapping similar to the one found in
one-electron systems. 
However, this two-electron trapping presents some new and interesting features.
Particularly remarkable is the presence of a hole (the empty state)
trapped in a superposition of the upper states. 
In the case where the two lower states are coupled by the electron-electron potential,
we propose a direct subnatural measure of these coupling coefficient.

Partially support by the Spanish Direcci\'{o}n General de Ense\~{n}anza
Superior e Investigaci\'{o}n Cient{\'{i}}fica 
(grants PB98-0935-C03-03, and PB98-0268), 
by the Direcci\'{o} Catalana de Recerca (contract 1999SGR00096), 
and by the Junta de Castilla y Le\'{o}n and Uni\'{o}n Europea, 
FSE (grant SA044/01), is acknowledged.
We thank G. Orriols and V. Ahufinger for critical reading of the manuscript.

\vskip2truecm

\begin{figure}[tbp]
\caption{Schematic representation of the single-electron states, 
 $\vert a \rangle$, $\vert b \rangle$, and $\vert c \rangle$, 
in a V configuration.}
\end{figure}
\vskip 2truecm

\begin{figure}[tbp]
\caption{Schematic representation of the two-electron states,
 $\vert A \rangle$, $\vert B \rangle$, and $\vert C \rangle$,
that automatically are arranged in a $\Lambda$ configuration.}
\end{figure}
\vskip 2truecm

\begin{figure}[tbp]
\caption{Representation of three level single-electron system. If only one
electron is considered (a) a uncoupled state $\vert d \rangle$ appears as a
superposition of the two excites states. State $\vert d \rangle$ may decay
to the lower state. If two electrons are considered (b) there is a hole that
can be trapped in this uncoupled state. This hole can not relax to any other
state }
\end{figure}
\vskip 2truecm

\begin{figure}[tbp]
\caption{Spectra of the dark state resonance and its two satellites induced
by the crossed correlations terms of the electron-electron interaction.
Figure represents the imaginary part of the $AC$ coherence 
 versus the detuning $\Delta_{\alpha}/ \gamma _{bc}$. 
It corresponds to 
$\gamma_{ac}=\gamma _{bc}$, 
$\alpha=\beta=0.1 \gamma _{bc}$, 
$\Delta_{\beta}=0$, and $\chi=0.3 \gamma _{bc}$. 
The three insets indicate the dresed-level couplings that contribute 
to each dark resonance}
\end{figure}

\end{mathletters}


\begin{references}

\bibitem{alzetta}  
G. Alzetta, A. Gozzini, L. Moi and G. Orriols, 
Nuovo Cimento B {\bf 36}, 5 (1976); 
E. Arimondo and G. Orriols, 
Lett. Nuovo Cimento {\bf 17}, 333 (1976); 
G. Alzetta, L. Moi and G. Orriols, 
Nuovo Cimento B {\bf 52}, 205 (1979).

\bibitem{orriols}   
G. Orriols, 
Nuovo Cimento B {\bf 53}, 1 (1979).

\bibitem{stroud}  
H. R. Gray, R. M. Whitley and C. R. Stroud, Jr., 
Opt. Lett. {\bf 3}, 218 (1978).

\bibitem{arimondo}  For a recent and complete review, see 
E. Arimondo, in Progress in Optics, edited by E. Wolf, 
(Elsevier, Amsterdam, 1996), Vol. {\bf 35}, p. 257.

\bibitem{harris}  
S. E. Harris, 
Physics Today, {\bf 50} 36 (1997)

\bibitem{mompart}  See, for example, 
J. Mompart, and R. Corbal\'an
J. Opt. B: Quantum Semiclass. Opt, {\bf 2}, R7 (2000)

\bibitem{grobe}  R. Grobe, S. L. Haan, and J. H. Eberly, 
Phys. Rev A {\bf 54}, 1516 (1996) 

\bibitem{orto-para}  
For analogy with the standard terminology with helium 
({\it ortho-helium} for the atom with parallel spins -triplet- and 
  {\it para-helium} for the atom with antiparallel spins -singlet-) 
we will refer throughout the paper to {\it ortho-atom} and {\it para-atom}, 
and to {\it ortho-system} and {\it para-system}.

\bibitem{rwa}  See, for example, 
Robert W Boyd, 
{\it Nonlinear Optics},
Academic Press, 1992.

\bibitem{angular}  
See, for example, 
A R Edmons, 
{\it Angular Momentum in Quantum Mechanics}, 
Princeton Univ. Press, 1974.

\end{references}
\end{document}